\documentclass[letterpaper]{article}

\begin{document}
\title{On 4D-Hawing Radiation from Effective Action}
\author{V. Mukhanov\thanks{
on leave from Institute for Nuclear Research, 117312 Moscow},
A. Wipf\\
Institut f\"ur Theoretische Physik,
Eidgen\"ossische Hochschule,\\
H\"onggerberg, Z\"urich CH-8093,
Switzerland\\
\\
and\\
\\A. Zelnikov\thanks{
on leave from: Lebedev Physical Institute, Moscow}
\\
Dept. of Physics, Univ. of Alberta,\\
Edmonton, AB T6G 2J1, Canada}
\date{ETH-TH/94-08}
\maketitle

\newcommand{\eqnn}[1]{\begin{eqnarray*}#1\end{eqnarray*}}
\newcommand{\eqnl}[2]{\par\parbox{11cm}
{\begin{eqnarray*}#1\end{eqnarray*}}\hfill
\parbox{1cm}{\begin{eqnarray}\label{#2}\end{eqnarray}}}
\newcommand{\eqnlb}[2]{\begin{equation}\fbox{$\displaystyle
#1 $}\label{#2}\end{equation}}
\newcommand{\eqngr}[2]{\par\parbox{11cm}
{\begin{eqnarray*}#1\\#2\end{eqnarray*}}\hfill
\parbox{1cm}{\begin{eqnarray}\end{eqnarray}}}
\newcommand{\eqngrlb}[3]{\par\parbox{11cm}
{\begin{eqnarray}\fbox{$\displaystyle   #1\\#2$}\end{eqnarray}}\hfill
\parbox{1cm}{\begin{eqnarray}\label{#3}\end{\eqnarray}}}

\newcommand{\eqngrl}[3]{\par\parbox{11cm}
{\begin{eqnarray*}#1\\#2\end{eqnarray*}}\hfill
\parbox{1cm}{\begin{eqnarray}\label{#3}\end{eqnarray}}}

\newcommand{\eqngrn}[3]{\begin{eqnarray*}#1 \\ #2\end{eqnarray*}}

\newcommand{\eqngrr}[3]{\par\parbox{11cm}
{\begin{eqnarray*}#1\\#2\\#3\end{eqnarray*}}\hfill
\parbox{1cm}{\begin{eqnarray}\end{eqnarray}}}

\newcommand{\eqngrrl}[4]{\par\parbox{11cm}
{\begin{eqnarray*}#1\\#2\\#3\end{eqnarray*}}\hfill
\parbox{1cm}{\begin{eqnarray}\label{#4}\end{eqnarray}}}

\newcommand{\refs}[1]{(\ref{#1})}

\def\mtx#1{\quad\hbox{{#1}}\quad}
\def\pa{\partial}
\def\gam{\gamma}
\def\egam{^{(\gam)}}
\def\es{\!=\!}
\def\gamr{\,^\gamma{\cal R}}
\def\omr{\,^\omega{\cal R}}
\def\lapgam{\triangle_\gamma}
\def\lapf{\triangle_f}
\def\lap{\triangle}
\def\olap{{1\ov\lap}}
\def\ov{\over}
\def\calr{{\cal R}}
\def\cd{{\cal D}}
\def\sqrg{\sqrt{g}}
\def\om{\omega}
\def\hb{\hbar}
\def\upf{^{(4)}\!}
\def\pamu{{\partial_\mu}}
\def\panu{{\partial_\nu}}
\def\al{\alpha}
\def\be{\beta}
\def\si{\sigma}
\def\gam{\gamma}
\def\pr{\prime}
\def\lam{\lambda}
\def\tr{\hbox{tr}\,}
\def\ta{{\tilde a}}
\def\tb{{\tilde b}}
\def\tc{{\tilde c}}
\def\tal{{\tilde\alpha}}
\def\tsi{{\tilde\sigma}}
\def\tbe{{\tilde\beta}}
\def\tga{{\tilde\gamma}}
\def\tde{{\tilde\delta}}
\def\ti{{\,\tilde i}}
\def\tj{{\,\tilde j}}
\def\hnn{\hat{0}}
\def\hr{\hat{R}}
\def\hi{\hat{I}}
\def\jm{\hat{N}}
\def\hio{\hat{I}_{\xi_1\lambda_1}}
\def\hit{\hat{I}_{\xi_2\lambda_2}}
\def\dxila{\delta_{\xi,\lambda}}
\def\dq{\dot{q}}
\def\cm{{\cal M}}
\def\cn{{\cal N}}
\def\ch{{\cal H}}
\def\cl{{\cal L}}
\def\ctal{{C_\tal}}
\def\ctbe{{C_\tbe}}
\def\ctga{C_\tga}
\def\tttr{t^\tga_{\tal \tbe}}
\def\tttw{t^\tbe_\tal}
\def\str{f^c_{ab}}
\def\ha{{1\over 2}}
\def\una{{\vec A}}
\def\unb{{\vec B}}
\def\und{{\vec D}}
\def\unpi{{\vec \pi}}
\def\unpa{{\vec\pa}}
\def\mtxt#1{\quad\hbox{#1}\quad}
\begin{abstract}
We determine the $s$-waves contribution
of a scalar field to the four dimensional effective action
for arbitrary spherically symmetric external
gravitational fields. The result is applied to $4d$-black
holes and it is shown that the energy momentum tensor
derived from the (nonlocal) effective action contains
the Hawking radiation. The luminosity is close to
the expected one in the $s$-channel. The energy momentum
tensor may be used as starting point to study the
backreaction problem.
\end{abstract}
\section{Introduction}
One of the interesting problems in quantum field theory in
curved space-times is to derive the induced energy momentum
tensor and in particular the Hawking radiation \cite{r1} from
the effective action approach \cite{r2,r3,r4}. The solution of this
problem is important for studying the backreaction
problem for black holes [5-8]. The effective action
for quantized matter fields in a black hole metric is
strongly nonlocal and should describe both the asymptotic Hawking
radiation and the vacuum polarization effects \cite{r9}.\par
Most of the recent works on the Hawking radiation and backreaction
problem (see, for instance [2,5-8]) concerned $2d$ black holes.
In particular, it has been shown that the $2d$ Hawking radiation
can be derived from the $2d$ effective action [2,9]. It is
a priori not clear whether these results are of relevance
for real $4d$ black holes.
On the other hand, the covariant perturbation theory for the $4d$ effective
action $\Gamma$ as developed in [3,10] seems to be very involved for
concrete calculations. The results obtained so
far are far from being complete.\par
In this paper we shall simplify the problem by considering
$s$-modes of minimally coupled massless scalar fields
propagating in an arbitrary spherically symmetric $4d$ spacetime.
We compute the contribution of these modes to the
$4d$ effective action $\Gamma$. The part of $\Gamma$
which is not invariant under Weyl-rescalings of the $(r,t)$-part
of the metric is exactly calculated. For the invariant part
an appropriate perturbation expansion is developed. As an
application the $s$-wave contribution to the Hawking flux
is obtained from the $s$-channel effective action for $4d$ black
holes. We demonstrate why and how the $2d$-calculations [2,5-8]
are relevant for realistic $4d$ black holes.\par
The calculations are performed in the Euclidean formalism.
The sign conventions, e.g.  for the Riemann tensor and
the signature when we return to Lorentzian space-time
after the calculations have been done, are the same
as in [11]. We set $c\es\hb\es G\es 1.$
\section{Setup}
The Euclidean action for the coupled gravitational and scalar fields is
\eqnl{
S_E=S_E^{grav}+S_E^{\phi}=-{1\ov 16\pi}\int \calr\sqrg\, d^4x
+\ha\int \sqrg g^{\al\beta}\pa_\al\phi\pa_\beta\phi\, d^4x.}{e1}
For spherically symmetric space-times it is convenient to choose adapted
coordinates for which the metric takes the form
\eqnl{
ds^2=g_{\mu\nu}dx^\mu dx^\nu=\gam_{ab}(x^a)dx^adx^b +\Omega^2(x^a)\;\omega_{ij}
dx^idx^j,}{e2}
where
\eqnl{
\omega_{ij} dx^idx^j=\big(d\theta^2+\sin^2\theta d\varphi^2\big)}{e3}
is the metric of $S^2$. The function $\Omega$ depends only on
the coordinates $(x^a)\es (x^0,x^1)\es (t,r)$ and $\gam_{ab}(x^a)$
is the metric in the $t-r$ sector. Note that $g^{ab}\es\gam^{ab}$.\par
In a spherically symmetric space-time we can expand a matter
field into spherical harmonics. In particular scalar fields in
the $s$-channel depend only on $t$ and $r$, $\phi=\phi(x^a)$.
For $s$-waves the action \refs{e1} reduces to the following $2d$ action
\eqngrl{
S_E=S_E^{grav}+S_E^\phi
&=&-{1\ov 4}\int \Big(\Omega^2 \gamr+\omr+2(\nabla \Omega)^2
\Big)\sqrt{\gam}\,d^2x}
{&+&2\pi\int \Omega^2(\nabla\phi)^2
\sqrt{\gam}\,d^2x,}{e4}
where we took into account that the volume of $S^2$ is equal to $4\pi$.
Here $\gamr$ is the scalar curvature of the $2d$-space
with metric $\gam_{ab}$, $\omr\es 2$ is the
scalar curvature of $S^2$ and $(\nabla\Omega)^2\es \gam^{ab}\pa_a\Omega
\pa_b\Omega$.\par
The purely gravitational part of the action \refs{e4} is almost the action
belonging to $2d$ dilatonic gravity with two exceptions:
first, the numerical coefficient in front of $(\nabla\Omega)^2$ is
different and second, the action \refs{e4} is not invariant under
Weyl transformations due to the $\omr$ term which
is the $2d$ analog of the cosmological constant in $4$ dimensions.
The action for the scalar field $\phi(t,r)$ is quite different
from the actions usually considered in $2d$-field theories [5-8]
because of the unusual coupling of $\phi$ to the dilaton field
$\Omega$. The action \refs{e4} is the $4d$-action for spherically symmetric
gravitational and scalar fields and as such should not be
regarded as just another $2d$-toy-model for gravity. \par
The independent field equations which follow from \refs{e4} are
\eqnl{
\lapgam\Omega-\Big({\gamr\ov 2}-4\pi(\nabla\phi)^2\Big)
\Omega=0}{e5}
from the variation of $\Omega$,
\eqnl{
\lapgam\Omega^2-\;\omr+8\pi\Omega^2(\nabla\phi)^2=0,}{e6}
which is the trace of the variation with respect to $\gam_{ab}$, and
the equation for the scalar field
\eqnl{
\nabla^a\big(\Omega^2\nabla_a\phi\big)=0.}{e7}
Here $\lapgam$ is the Laplace-Beltrami operator
in $2$-dimensional space-time with the metric
$\gam_{ab}$. Note that the matter part of the action (4) is invariant
under 2-dimensional Weyl transformations, $\gam_{ab}\to e^{2\sigma}\gam_{ab}$,
and hence the partial trace $T^a_{\;a}$ of the energy momentum tensor
vanishes for spherically symmetric scalar fields.\par
Without matter ($\phi\es 0$) the $2d$-Euclidean black holes
\eqnl{
^{(2)}ds^2=\Big(1-{r_g\ov r}\Big)dt^2+{dr^2\ov 1-r_g/r},
\qquad\quad\Omega=r}{e8}
are solutions of (\ref{e5},\ref{e6}) as it should be.
\section{Effective Action}
In this section we determine the $s$-wave contribution of the
quantized scalar field to the effective action. In particular we
shall show how this problem can be reduced to a $2d$ problem.
Then we shall calculate the non Weyl-invariant part of the
$s$-channel effective action exactly and develop a perturbation
theory for the Weyl-invariant part.\par
The Euclidean $4d$-effective action $\Gamma$ is defined as
\eqnl{
e^{-\Gamma}=\int \cd \phi\,e^{-S_E^\phi},}{e9}
where
\eqnl{
S_E^\phi=-\ha\int\phi\triangle\phi\,\sqrt{g}\;d^4 x}{e10}
is the Euclidean action for the minimally coupled scalar field
and $\triangle\es\triangle_g$ is the $4d$-Laplace-Beltrami operator.
To define the (formal) diffeomorphism invariant measure in the path integral
\refs{e9} we expand the field $\phi(x^\al)$ in terms of the eigenfunctions
of $-\triangle$. For a spherically symmetric space-time and adapted
coordinates \refs{e2} this expansion reads
\eqnl{
\phi(x^\al)=\sum_{nlm}\phi_{nlm},}{e10a}
where
\eqnl{
\phi_{nlm}=\phi^{lm}_n(t,r) Y_{lm}(\theta,\phi),\qquad
-\triangle \phi_{nlm}=\lam_{nl}\phi_{nlm}.}{e11}
Here the $Y_{lm}$ are the spherical harmonics and the eigenmodes
are normalized with respect to the $4$-metric:
\eqnl{
\langle \phi_{nlm}\vert\phi_{n^\pr l^\pr m^\pr}\rangle=
\int \phi_{nlm}\phi_{n^\pr l^\pr m^\pr}\sqrg\;d^4x=
\delta_{nn^\pr}\delta_{ll^\pr}\delta_{mm^\pr}.}{e12}
Then the path integral becomes
\eqnl{
e^{-\Gamma}=\int\prod_{nlm}dc_{nlm}\exp\Big(-\ha\sum_{nlm}\lam_{nl}\,
c^2_{nlm}\Big)=\exp\Big(-\sum_l (2l+1)\Gamma_l\Big),}{e13}
where
\eqnl{
e^{-\Gamma_l}=\int\prod_n dc_{nlm}\exp\Big(-\ha\sum_n\lam_{nl}\,
c^2_{nlm}\Big)}{e14}
is the contribution of the modes with quantum numbers $(l,m)$
and we took into account that the eigenvalues do not depend on the magnetic
quantum number $m$.\par
The integrals (\ref{e13},\ref{e14}) are of course ultraviolet divergent and
must be regularized. We shall use the $zeta$-function regularization
[12]. Any other covariant regularization of \refs{e13} would yield the same
result up to integrals of local terms of the form $\sqrg$, $\sqrg\calr$ and
$\sqrg\calr^2,\sqrg\pa\pa\calr$ [13]. The coefficients
of these ambigues terms should be determined by experiments or observations
in any case. One can regularize every $\Gamma_l$ separately
and then sum over all angular momenta to recover the total
effective action $\Gamma\es\sum_l (2l+1)\Gamma_l$. In general
the sum of the regularized $\Gamma_l$'s is still quadratically
divergent. However, the remaining quadratic and logarithmic divergences
can be absorbed by redefining coefficients of the local counterterms.
For the finite non-local terms of interest the regularization
commutes with taking the sum over the angular momentum
sectors. Hence, to obtain the nonlocal contribution of the
different sectors to $\Gamma$ one can apply the $\zeta$-function
regularization for every sector separately.\par
In this paper we shall calculate only the contribution $\Gamma_s\equiv
\Gamma_{0}$ of the $s$-wave scalar fields to the total $4d$ effective
action:
\eqnl{
e^{-\Gamma_s}=\int \prod_n dc_{n00}\exp\Big(-\ha\sum_n\lam_{n0}\,
c_{n00}^2\Big).}{e15}
Now we shall show how \refs{e15} relates to the effective action
of a $2$-dimensional theory. For that we introduce the complete
set of rescaled $s$-modes
\eqnl{
\varphi_n(t,r)=\sqrt{4\pi}\,\Omega(t,r)\phi_{n00}(t,r)}{16}
which are orthonormal with respect to the $2$-metric $\gamma_{ab}$
\eqnl{
\langle \varphi_n\vert \varphi_{n^\pr}\rangle
=\int\;\varphi_n\varphi_{n^\pr}\sqrt{\gam}\,d^2x=\delta_{nn^\pr},}{e17}
contrary to the $\phi_{n00}$, which are orthonormal with respect
to the $4$-metric $g_{\al\beta}$. Then any field
$\varphi(t,r)\es \sqrt{4\pi}\Omega\phi_{l=0}(t,r)$ can be expanded as
\eqnl{
\varphi=\sum_n\;c_n\varphi_n.}{e18}
Also note that since $\Omega$ in \refs{e2} had the dimension of a length
(if we keep the dimension of the gravitational constant $G$),
$\varphi$ becomes dimensionless as required for a $2$-dimensional
scalar field. \par
It is easy to see that the $\varphi_n$ are the eigenmodes
of the $2d$-operator
\eqnl{
\hat O=\Big(-\lapgam+{\lapgam\Omega\ov\Omega}\Big)}{e19}
with the same eigenvalues $\lam_n\es \lam_{n0}$ an in \refs{e11}.
Then \refs{e15} can be rewritten as the functional integral
of a field theory in $2$-dimensional space time with the
metric $\gamma_{ab}$:
\eqnl{
e^{-\Gamma_s}=\int \cd\varphi \exp\Big(-\ha\int \varphi\hat O
\varphi\,\sqrt{\gam}\,d^2x\Big),}{e20}
where the measure $\cd \varphi$ is the usual (formal) Lebeques measure.
The classical action in \refs{e20} is of course just the action $S_E^\phi$
in \refs{e4} rewritten in terms of the  $2d$-scalar field $\varphi$.
{}From \refs{e20} it follows at once that
\eqnl{
\Gamma_s=\ha\log\det \hat O.}{e21}
Thus, calculating the $s$-waves contribution to the effective
action reduces to the problem of calculating the determinant
of the operator $\hat O$ defined in $2d$-space with metric
$\gamma_{ab}$. This operator has the nice and important property
that it transforms homogeneously under Weyl-rescalings
\eqnl{
\gam_{ab}\to e^{2\sigma}\gam_{ab}
\Longrightarrow
\hat O\to e^{-2\sigma}\hat O.}{e22}
This immediately implies that the classical $2d$-action for $\varphi$
is Weyl invariant. As it is well-known [14], the $2d$-effective action
ceases to be Weyl invariant and the breaking is determined
by the trace anomaly which is proportional to the first Seeley-deWitt
coefficient $a_1$, which in our case is
\eqnl{
a_1={\gamr\ov 6}-{\lapgam\Omega\ov\Omega}.}{e23}
Thus we have
\eqnl{
\gamma^{ab}{2\ov\sqrt{\gam}}{\delta\Gamma_s\ov\delta\gam^{ab}}
={a_1\ov 4\pi}.}{e24}
This equation can easily be integrated if we choose isothermal
coordinates (the conformal gauge)
\eqnl{
\gam_{ab}=e^{2\sigma}\gam^f_{ab},}{e26}
where $\gam^f_{ab}$ is the metric of the flat $2d$ space. In this
gauge
\eqnl{
\gamr=-2\lapgam\sigma=-2 e^{-2\sigma}\lapf\sigma}{e27}
and \refs{e24} simplifies to
\eqnl{
{\delta \Gamma_s\ov \delta \sigma}={1\ov 4\pi}\Big({1\ov 3}\lapf
\sigma+{\lapf \Omega\ov \Omega}\Big),}{e28}
where $\lapf $ is the Laplace operator on flat space which does
not depend on $\sigma$. Integrating \refs{e28} and expressing $\sigma$
in terms of $\gamr$ by means of eq. \refs{e27} one ends up
with
\eqngrl{^{(n)}\Gamma_s[\sigma,\Omega] &\equiv&
\Gamma_s[\sigma,\Omega]-\Gamma_s[\sigma=0,\Omega]}
{&=&{1\ov 8\pi}\int\Big[{1\ov 12}\gamr{1\ov\lapgam}\gamr
-{\lapgam\Omega\ov\Omega}{1\ov\lapgam}\gamr\Big]\sqrt{\gam}\,
d^2x}{e29}
which is manifestly invariant under $2d$-coordinate transformations.
The first term on the r.h.s in \refs{e29} has been
obtained previously in [2,9].In [2] it has been shown that it
leads to the Hawking
radiation for $2d$ black holes. We shall see in the following
section how it is related with the $s$-wave radiation of $4d$ black holes.
However, there is the second term which, as can easily be checked,
yields a infalling radiation flux. The amplitude of this
radiation exceeds the outgoing flux coming from the first
term by a factor $6$. However, we must not forget that in \refs{e29}
we calculated only that part of $\Gamma_s$ which is non-invariant
under $2d$-Weyl transformations. The total $s$-wave effective action is
\eqnl{
\Gamma_s=^{(n)}\!\Gamma_s+^{(i)}\!\Gamma_s,}{e30}
where
\eqnl{
^{(i)}\Gamma_s=\Gamma_s[\sigma=0,\Omega]=\ha\log\det\Big(-\lapf +
{\lapf \Omega\ov\Omega}\Big)\equiv
\ha\log\det\hat O_f}{e31}
is the part of the total effective action which is invariant under
$2d$-Weyl transformations. To get the complete result we should
also calculate the determinant of $\hat O_f$ on flat space and then
restore the metric $\gam_{ab}$ in the obtained expression such
as to recover general covariance. \par
Unfortunately $\log\det\hat O_f$ cannot be calculated exactly and
we must resort to some perturbation expansion. The covariant
perturbation theory developed in [3,10] seems to be of no help
here because of severe infrared divergences. These are
related with the non-analyticity of the effective action
in the potential
\eqnl{
V^f={\lapf \Omega\ov \Omega}.}{e32}
Instead we write the heat kernel for $\hat O_f$ in the form
\eqnl{
K(x,x;\tau)={\mu^2\ov 4\pi\tau}\exp\Big(-{\tau\ov \mu^2}W^f(x;\tau)\Big)
}{e33}
and develop the perturbation theory for $W$ in powers of the potential
$V$ (see appendix B). The arbitrary mass $\mu$ has been introduced
such that $\tau$ becomes dimensionless.

Using the $\zeta$-function regularization (see appendix A) we immediately
arrive at the following finite expression for the effective
action in terms of $W$ and $V$:
\eqngrrl{
^{(i)}\Gamma_s&=&\Gamma_s^{CW}+\Gamma_s^{BS},\mtxt{where}}
{\Gamma_s^{CW}&=&{1\ov 8\pi}\int\Big\{V^f-V^f\log{V^f\ov\mu^2}\Big\}
\sqrt{\gam^f}d^2x}
{\Gamma_s^{BS}&=&\int_0^\infty d\tau \Big\{{W^f-V^f\ov \tau W^f}(\tau W^f)^\pr
\exp(-{\tau\ov\mu^2}W^f)\Big\}\sqrt{\gam^f}d^2x}{e34}
and the prime means differentiation with respect to $\tau$.
$\Gamma_s^{CW}$  correspond to the $2d$ Coleman-Weinberg
potential [16]. For constant $V^f$ we have $W^f=V^f$ so that
$\Gamma_s^{BS}$ vanishes and the Coleman-Weinberg
potential is the exact result. In this section we will neglect
$\Gamma_s^{BS}$ in \refs{e34} which is proportional to
the derivatives of the potential $V$. This approximation corresponds
to the simple classical approximation to the heat kernel \refs{e33}
and yields the $4d$ $s$-channel Hawking radiation without
backscattering effects. The more involved problem of including
the backscattering will be discussed in section 5.

Next we need to covariantize the Weyl-invariant Coleman-Weinberg
contribution to \refs{e34}, that is restore the original metric $\gam_{ab}$.
Taking into account that
\eqnl{
V^f\equiv {\lap_f\Omega\ov \Omega}=e^{2\sigma}{\lapgam\Omega\ov\Omega}}{e34a}
and expressing $\sigma$ in terms of of $\calr$ via \refs{e27}
we obtain the following $2d$-covariant result
\eqnl{
\Gamma_s^{CW}={1\ov 8\pi}\int{\lapgam\Omega\ov\Omega}\Big(
1-\log{1\ov\mu^2}{\lapgam\Omega\ov\Omega}+{1\ov\lapgam}\gamr\Big)
\sqrt{\gam}\,d^2x.}{e35}
Note that only the last term is nonlocal and contributes to the
Hawking flux. The action $\Gamma_s^{CW}$
is invariant under Weyl-transformations \refs{e26} as required and hence
does not contribute to the trace of the $2d$
energy momentum tensor (EMT). In the next section we shall
see that it also does not contribute to the partial trace
$T^a_{\;a}$ of the $4d$ EMT. However, it contributes
to the total trace $T^\al_{\;\al}$.
The free mass-parameter $\mu$ corresponds to the renormalization
arbitrariness.\par
Combining \refs{e29} and \refs{e35} one finally obtains for the $s$-channel
effective action in the no-backscattering approximation
\eqnl{
^{(n)}\Gamma_s+\Gamma_s^{CW}=
{1\ov 8\pi}\int \Big({1\ov 12}\gamr{1\ov\lapgam}\gamr
-{\lapgam\Omega\ov\Omega}\big(1+\log{\lapgam\Omega\ov\mu^2\Omega}
\big)\Big)\sqrt{\gam}\,d^2x.}{e36}
We see that the nonlocal term in \refs{e35} cancels against the second
term in \refs{e29} which yields a negative contribution to the
Hawking flux. However, in the section 5 we shall see
that the region near a black hole contributes significantly
to the (so far neglected) $\Gamma_s^{BS}$ in \refs{e34} and effectively reduces
the coefficient in front of the nonlocal term in \refs{e36}.
Physically this corresponds to a decreasing of the Hawking
flux due to backscattering effects.
\section{$4d$-Energy Momentum Tensor and Hawking Radiation}
The $4d$-EMT can be derived from the $4d$ effective action \refs{e9}
according to
\eqnl{
T_{\al\beta}={2\ov\sqrg}{\delta\Gamma\ov\delta g^{\al\beta}}.}{e37}
The $s$-waves contribution to $T_{\al\beta}$ is then gotten by
inserting $\Gamma_s$ for $\Gamma$ in \refs{e37}. For that we rewrite
the $s$-channel effective action \refs{e21} in terms of the $4d$ metric
as
\eqnl{
\Gamma_s={1\ov 4\pi}\int {1\ov \Omega^2}\,\cl_s\sqrg\,d^4x.}{e38}
Then using the symmetry properties and taking into account that
$g_{\al\beta}\es (\gam_{ab},\Omega^2\om_{ij})$ one obtains the
following formulae for the non-vanishing components of the $4d$
EMT:
\eqnl{
T^a_{\;b}={1\ov 2\pi \Omega^2}{1\ov\sqrt{\gam}}\gam^{ac}
{\delta\Gamma_s\ov\delta\gam^{cb}}\mtxt{,}
T^i_{\;j}=-{1\ov 8\pi\Omega}{1\ov \sqrt{\gam}}{\delta\Gamma_s\ov
\delta\Omega}\delta^i_{\;j}.}{e39}
Without backscattering effects $\Gamma_s$ is given by \refs{e36} and
the functional derivative is to be calculated in $2d$-space. Straightforward
calculations lead to the following explicit expressions for the EMT
($\lap=\lapgam,\calr=\calr^\gamma$):
\eqngrl{
T^a_{\;b}&=&{1\ov 4\pi\Omega^2}{1\ov 48\pi}
\Big[-2\nabla^a\nabla_b(\olap\calr)+\nabla^a(\olap\calr)\nabla_b
\olap\calr}
{&&\qquad\quad+\delta^a_{\;b}\Big(2\calr-\ha\nabla^c(\olap\calr)
\nabla_c(\olap\calr)\Big)\Big]+\hbox{local terms}}{e40}
The components $T^i_{\;\,j}$ contain only local terms which
give rise to vacuum polarization effect. Here we are mainly
interested in particle creation and for that reason
skipped all local terms in \refs{e40}.
We stress that up to this point our results apply to arbitrary
spherically symmetric backgrounds. Thus \refs{e40} describes the
$s$-channel particle creation (and vacuum polarization) for minimally
coupled scalars propagating in an {\em arbitrary spherically symmetric}
spacetime,
e.g. of a collapsing star. Here we are mainly concerned with
the Hawking radiation and leave other interesting applications
to a forthcoming publication.\par
To get the flux of the Hawking radiation we need to go back
to Lorentzian space-time by changing the signs in the
appropriate places. According to the results in [2,17]
we arrive at the in-vacuum EMT by  replacing $-1/\lap$ by
the retarded Greens function $G^-$. Only the second term in
\refs{e40} contributes to the Hawking radiation. The calculations
leading to the corresponding flux are analogous to the ones which have
been done by Frolov and Vilkovisky [2]. Thus we may skip them here
by referring the reader to that paper. \par
The luminosity of the black hole, which is obtained from the
Lorentzian (in-vacuum) version of the EMT \refs{e40} is then found to be
\eqnl{
L=-{\pi\ov 12}{1\ov (8\pi M)^2},}{e42}
where $M$ is the mass of the black hole. This exactly coincides
with the total $s$-waves flux of the Hawking radiation obtained
by standard methods [14] without taking backscattering effects
into account.
\section{Backscattering effect}
To take into account the backscattering of Hawking radiation
we must calculate $\Gamma_s^{BS}$ in \refs{e34},
\eqnl{
\Gamma_s^{BS}={1\ov 8\pi}\int I\sqrt{\gam^f}d^2x\mtxt{,}
I=\int d\tau {W^f-V^f\ov \tau W^f}\big(W^f\tau\big)^\pr\,e^{-W^f\tau},}{e43}
to the effective action. For that we develop the perturbation
expansion for $W^f$ in powers of the potential $V^f$. This
expansion in presented in appendix B. Up to linear order
in $V^f$ we find
\eqnl{
W^f(x;\tau)=\sum_{n=0}^\infty {n!\ov (2n+1)!}\big({\tau\lap\ov
\mu^2}\big)^n \,V^f(x)+O(\nabla V^f\cdot \nabla V^f).}{e44}
To simplify the analysis we choose
the natural radial variable introduced by Regge and Wheeler
\eqnl{
r^*=r+2M\log\vert{r\ov 2M}-1\vert}{e46}
so that the $(r,t)$-part of the (Euclidean) black hole metric
takes the form
\eqnl{
ds^2=\big(1-{2M\ov r}\big)\big(dt^2+dr^{*2}\big).}{e45}
Note that $r^*\to\infty$ as $r\to\infty$, but also
$r^*\to-\infty$ as $r\to 2M$.
In this coordinate system  the potential $V^f$
reads
\eqnl{V^f(r^*)=e^{2\sigma}{\lapgam\Omega\ov\Omega}=\big(1-{2M\ov
r}\big){2M\ov r^3},}{e48}
where $r$ should be expressed in terms of $r^*$ via \refs{e46}.
Since the potential $V^f$ depends only on one coordinate, namely $r^*$,
the asymptotic series \refs{e44} can be converted into an
integral
\eqnl{
W^f(r^*;\tau)=\sqrt{\pi\mu^2\ov 4\tau}\int_{-\infty}^\infty
V(\tilde r^*)\Bigg(1-\Phi({\mu\vert \tilde r^*-r^*\vert\ov\sqrt{\tau}})
\Bigg)d\tilde r^*,}{e49}
where
\eqnl{
\Phi(x)=\int_0^x e^{-t^2}dt}{e50}
is the error function. For $r^*\gg 2M$, where $r^*\approx r$, we
can find a good approximation to the integral \refs{e49}
for different values of $\tau/\mu^2$.  The $\tau/ \mu^2$-dependence
of $W^f$ for $r\gg 2M$ is depicted in figure 1. We see that for
$\tau/\mu^2\ll V^f/\lapf V^f$ the function $W^f$ slowly increases
as a function of $\tau$, starting with $W^f(r;\tau\es 0)=V^f(r)$.
Then, in a very short interval in the vicinity of $\tau/\mu^2\sim
V^f/\lapf V^f$ it increases dramatically from $M/r^3$ to
$1/Mr$. When $\tau$ is much bigger
then $W^f$  decreases as $1/\sqrt{\tau}$. Since
$W^f\tau\sim\sqrt{\tau}$ when $\tau\to\infty$ the expression for
the effective action is infrared-convergent.\par
Clearly, the small interval in the vicinity of $V^f/\lapf V^f$,
where $W^f$ changes a lot, gives the main contribution to the
integral in \refs{e34}. In this interval we have $W^f\gg V^f$,
$(W^f)^\pr\tau \gg W^f$ and we can estimate the integral \refs{e43}
as
\eqnl{
I\sim\int_0^\infty d\tau\,(W^f)^\pr\exp\big(-{W^f\tau\ov\mu^2}\big)
\sim W^f(\tau_0),}{e51}
where $\tau_0$ is the value of $\tau$ for which
$\tau_0 W^f(\tau_0)\sim\mu^2$.
For potentials for which $W^f$ has the qualitative shape depicted
in figure 1, we have $\tau_0/\mu^2\sim V^f/\lapf V^f$. Thus one
obtains
\eqnl{
I\sim \xi {\lapf V^f\ov V^f},}{e52}
where $\xi$ is some fudge coefficient. In our approximate
treatment of the integral in \refs{e43} we cannot get the
exact value for this coefficient. Note that at $\tau\sim\tau_0$
the main contribution to the integral \refs{e49} which
defines $W^f$ comes from the region near the black hole
horizon. This confirms that \refs{e52} actually takes into account
the backscattering of the Hawking radiation in the potential
of the black hole, which is most effective near a black hole.\par
Now we need to restore the metric $\gam_{ab}$ in \refs{e52}.
Taking into account eqs. \refs{e27} and \refs{e35} we find that
\refs{e52} leads to the following contribution to the total
effective action
\eqnl{
\Gamma_s^{BS} =-{\xi\ov 8\pi}\int\Big(\gamr{1\ov\lapgam}\gamr
+\hbox{local terms}\Big)\sqrt{\gam}d^2x.}{e53}
This must be added to \refs{e36} to get the $s$-channel
effective action. Notice that the
nonlocal term in \refs{e53} cancels part of the nonlocal term in \refs{e36}
and diminishes the total Hawking flux. Comparing our result
with that obtained by other means [] we conclude that $\xi$
should be about $10$ percent less than $1/12$.

\section{Conclusions}
We have calculated the contribution of the $s$-waves of massless
minimally coupled scalars to the $4d$-effective action in an
arbitrary spherically symmetric external gravitational field. The problem
was to a large extend simplified by reducing the $s$-waves sector
to an effective $2$-dimensional, classically Weyl-invariant theory.
Of course, it is obvious that the $s$-wave channel reduces
to a $2$-dimensional theory. But during the reduction process
one needs to rescale the spherically symmetric scalar field
(see (e16)) such that the new measure
in the path integral belongs to a scalar field propagating in
a $2$-dimensional spacetime. The field theory for the $2$-dimensional
field is a conformal field theory. This observations permitted
us to calculate the Weyl non-invariant part of the effective action
exactly.
Then the problem reduces to the calculation of the
$2d$-Weyl invariant part
which actually is an effective action in
$2d$ flat spacetime. To calculate this we developed
the perturbation expansion which works well in the
case of black holes and permits us to take into account
the backscattering of the Hawking radiation by the
gravitational field of the black hole. As an application we
derived the explicit form of that part of the stress-energy tensor
which leads to the Hawking radiation.\par
However, the range of applicability of our main results
is not at all restricted to the black hole
physics. They hold for arbitrary spherically symmetric backgrounds and
consequently can be applied to study collapse problems, e.g. the
particle production by time-dependent spherical gravitational fields.
\paragraph{Acknowledgments:}
We thank A. Barvinsky, V. Frolov, I. Sachs and C. Schmid for illuminating
discussions. This work has been supported in part by
the Swiss National Science Foundation.
\appendix
\section{Regularization of the effective action and infrared problem}
The effective action $\Gamma$ expressed in terms of the heat kernel
is divergent. Since we are only interested in the finite part
of $\Gamma$ rather than the divergent one we derive here the
expression for the $\zeta$-function
regularized $\Gamma$ in terms of some function
$W(x;\tau)\equiv W(x,x;\tau)$ which is defined via the
heat kernel $K(x,y,\tau)$ in $d$-dimensional space as
\eqnl{
K(x,y,\tau)={\mu^d\ov
(4\pi\tau)^{d\ov 2}}\exp\Big[-{\mu^2(x-y)^2\ov 4\tau}-{W(x,y;\tau)\,\tau\ov
\mu^2}\Big].}{a1}
Here $\mu$ is an arbitrary mass-parameter introduced for
dimensional reasons. Since this parameter can be easily
restored in the final results we will set it to one
to simplify the formulae. We will see how one solves the infrared
problem which is usually met when one uses the Seeley-deWitt
expansion for the heat kernel in the massless case. Keeping
in mind the other possible applications besides the black
hole physics we consider the general $d$-dimensional case.
The formula \refs{a12} below which we need for our purposes
is then gotten as particular example. Finally we show how
to construct the asymptotic series for the finite part
of $\Gamma$ in terms of the Seeley-deWitt coefficients.\par
An efficient methods (which respects the diffeomorphism
invariance) to calculate the effective action is the
$\zeta$-function regularization [12] in terms of which
\eqnl{
\Gamma=-\ha{d\zeta (s)\ov ds}\vert_{s=0}.}{a2}
Here $\zeta(s)$ is the meromorphic function which for
$s>d/2$ has the integral representation
\eqnl{
\zeta(s)={1\ov\Gamma(s)}\int_0^\infty d\tau \tau^{s-1}\tr K(\tau)}{a3}
in terms of the heat kernel \refs{a1}. First, let us introduce
instead of $\tau$ the new variable
\eqnl{\eta=W(x;\tau)\tau}{a4}
assuming that $W(x;\tau)\tau$ increases monotonically from
$0$ to $\infty$ when $\tau$ runs through the same interval.
Then the expression \refs{a3} takes the form
\eqnl{
\zeta(s)={1\ov (4\pi)^{d\ov 2}\Gamma(s)}\tr_x\int_0^\infty
d\eta \,e^{-\eta}\eta^{s-1-{d\ov 2}}W^{{d\ov 2}-s}\Big(
1-{\eta W^\pr\ov W}\Big),}{a5}
where now prime means differentiation with respect to $\eta$. The integral
in \refs{a5} is convergent for $s>d/2$. Integrating sufficiently
often by parts one gets the following expression for
the analytic continuation of $\zeta(s)$ to $s\to 0$:
\eqnl{
\zeta(s)=\big({-1\ov 4\pi}\big)^{d\ov 2}
{\Gamma(s-{d\ov 2})\ov\Gamma(s)\Gamma(s+1)}
\int_0^\infty \eta^s\Big(W^{{d\ov 2}-s}\,e^{-\eta}\Big)^{({d\ov 2})}d\eta}{a6}
in $d=2,4,\dots$ dimensions and
\eqnl{
\zeta(s)={1\ov 2\sqrt{\pi}}\big({-1\ov  4\pi}\big)^{d\ov 2}
{\Gamma(s-{d\ov 2})\ov\Gamma(s)\Gamma(s+\ha)}
\int_0^\infty \eta^{s-\ha}\big(W^{{d\ov 2}-s}\,e^{-\eta}\big)^{({d\ov
2}-\ha)}d\eta}{a7}
in $d=1,3,\dots$ dimensions. Here $(\dots)^{(n)}$ denotes the
$n$'t derivative with respect to $\eta$. Calculating the $s$-derivative
at $s=0$ we arrive at the following formulae for the finite parts
of the effective actions
\eqngrrl{
\Gamma&=&\ha{1\ov (4\pi)^{d\ov 2}}{1\ov \Gamma({d\ov 2}+1)}
\sum_{k=0}^{{d\ov 2}-1}(-1)^{{d\ov 2}-1-k}{{d\ov 2}-1\choose k}}
{&&\cdot\tr_x\Bigg\{-(W^{d\ov 2}\log
W)_0^{(k)}+(\sum_{n=1}^{d/2}\;{1\ov n})(W^{d\ov 2})_0^{(k)}}
{&&+
\int_0^\infty d\eta\,{e^{-\eta}\ov\eta}\Big[(W^{d\ov 2})^{(k)}-
(W^{d\ov 2})^{(k)}_0\Big]\Bigg\}}{a8}
in even dimensions and
\eqngrl{
\Gamma&=&\ha{\sqrt{\pi}\ov (4\pi)^{d\ov 2}}{1\ov \Gamma({d\ov 2}+1)}
\sum_{k=0}^{{d-1\ov 2}}(-1)^{{d\ov 2}-\ha-k}
{{d\ov 2}-\ha\choose k}}
{&&\qquad\quad\cdot \tr_x
\int_0^\infty d\eta {e^{-\eta}\ov \sqrt{\eta}}(W^{d\ov 2})^{(k)}}{a9}
in odd dimensions, where the subscript $0$ means that the
derivative should be taken at $\eta=0$.
Taking into account \refs{a4} we can see that the integrals
in (\ref{a8},\ref{a9}) are convergent both in the ultraviolet
($\eta\to 0$) and infrared ($\eta\to\infty$) regions. Note that even
for massless fields no infrared divergences appear. Of course
we assumed that the map $[0,\infty)\ni\tau\to\eta$ is bijective
which in particular implies that $W(x,\tau)$ does not
decay faster than $1/\tau$ for large $\tau$. Now we shall show
how to relate the finite part of the effective action to the
Seeley-deWitt coefficients.\par
In $2$-dimensions the formulae simplify considerably and
the effective action \refs{a8} reads
\eqnl{
\Gamma={1\ov 8\pi}\tr_x\Big[W_0-W_0\log W_0+
\int_0^\infty d\eta {e^{-\eta}\ov \eta}\big(W-W_0\big)\Big].}{a10}
For an operator
\eqnl{\hat O=-\lap+V(x)}{a11}
in flat space we have $W_0=V$. The first two terms in \refs{a10}
correspond then to the $2d$-Coleman-Weinberg potential and the integral
gives the correction which vanishes for constant $V$.
Expanding $W(x;\eta)$ in a Taylor series and integrating over $\eta$
we find the following asymptotic series for $\Gamma$:
\eqnl{
\Gamma={1\ov 8\pi}\tr_x\Big(W_0-W_0\log W_0+\sum_{n=1}^\infty
{1\ov n}W_0^{(n)}\Big).}{a12}
Note that the derivative with respect to $\eta$ is related to
the $\tau$-derivative via
\eqnl{
{\pa\ov\pa\eta}={1\ov W+{\pa W\ov\pa\tau}\tau}{\pa\ov\pa\tau}.}{a13}
Thus, the series \refs{a12} can be viewed as an expansion of the
effective action in terms of the $\tau$-derivatives of $W(x;\tau)$.
In particular, the first few terms in \refs{a12} can be
explicitly written as
\eqngrl{
\Gamma&=&{1\ov 8\pi}\tr_x\Bigg(W_0-W_0\log W_0+{1\ov
W_0}({\pa W\ov\pa\tau})_0+\ha{1\ov W^2_0}({\pa^2 W\ov\pa\tau^2})_0}
{&& -{2\ov W^3_0}\big({\pa W\ov\pa\tau}\big)^2_0\Bigg)
+O\Big({1\ov W^3_0}({\pa^3 W\ov \pa\tau^3})_0,{1\ov W^4_0}
\big({\pa W\ov\pa\tau}\big)^3_0,\dots\Big)}{a14}
Expanding \refs{a1} in powers of $\tau$ and comparing it with the
Seeley-deWitt expansion for the heat kernel on the diagonal
\eqnl{
K(x,x;\tau)={1\ov (4\pi\tau)^{d\ov 2}}\sum a_n\tau^n}{a15}
we can express the derivatives of $W$ at $\tau=0$ in terms of
the Seeley-deWitt coefficients as
\eqngrrl{
W_0&=& -a_1}
{({\pa W\ov\pa\tau})_0&=&\ha a_1^2-a_2}
{({\pa^2 W\ov \pa\tau^2})_0&=&2a_1a_2-2a_3-{2\ov 3}a_1^3,\dots}{a16}
Then the effective action \refs{a14} can be rewritten in terms of
the $a_n$ as
\eqngrl{
\Gamma&=&{1\ov 8\pi}\tr_x\Big\{a_1-a_1\log a_1+({a_2\ov a_1}-\ha a_1)}
{&&\qquad (-{a_3\ov a_1^2}+{a_2^2\ov a_1^3}-{1\ov 12}a_1)+\dots\Big\}.}{a17}
In particular, for operators of the form \refs{a12} in a $2d$ flat
spacetime one gets
\eqngrl{
\Gamma&=&{1\ov 8\pi}\tr_x\Bigg\{V-V\log V+{1\ov 6}{\lap V\ov V}}
{&&\qquad -{1\ov 12}{(\nabla V)^2\ov V^2}+{1\ov 60}{\lap^2 V\ov V^2}
-{1\ov 36}{(\lap V)^2\ov V^3}+\dots\Bigg\}.}{a18}
Note that the asymptotic expansion \refs{a12} (and correspondingly
\refs{a18}) is good only if the potential $V$ is big compared
with its derivatives. In this case the formal expansion parameter is
$\lap V/V^2\ll 1$. When this condition is not met, as for instance
for the black hole metric,  then we must
work directly with \refs{a10}.
\section{Perturbation theory for $W$}
To calculate the effective action for an operator \refs{a11}
in flat spacetime we need to develop some perturbation expansion
for the heat kernel $\hat K(\tau)$ which satisfies
\eqnl{
{\pa \hat K\ov\pa\tau}=-\hat O\hat K\mtxt{and}
\hat K(\tau=0)=\hat 1.}{b2}
In the coordinate representation we write the heat kernel in the
form \refs{a1} (again we set $\mu=1$)
and derive the perturbation series for $W$ in powers of the potential
$V$. Keeping in mind other possible applications of the perturbation
expansion (e.g. in statistical mechanics [15]) we consider
an arbitrary number of dimensions $d$. Substituting \refs{a1}
into \refs{b2} we obtain the following equation for $W$:
\eqnl{
\tau{\pa W\ov\pa\tau}=\lap W\tau-(x-y)^i\nabla_iW-(\nabla W)^2\tau^2+V-W.}{b5}
Making the 'ansatz'
\eqnl{
W(x,y;\tau)=\sum b_n(x,y)\tau^n}{b6}
we immediately arrive at the recurrence relations for the $b_n$:
\eqngrl{
b_0+(x-y)^i\nabla_i b_o&=&V,}
{2b_1+(x-y)^i\nabla_i b_1&=&\lap b_0\;,}{b7}
and for $n>2$
\eqnl{
(n+1)b_n+(x-y)^i\nabla_i b_n=\lap b_{n-1}-\sum_{p=0}^{n-2}
\nabla_ib_p\nabla^ib_{n-p-2}.}{b8}
For computing the effective action or partition function
it suffices to know $K$ and correspondingly the $b_n$ on the diagonal
$x=y$. Taking this coincidence limit in (\ref{b7},\ref{b8})
(of course, after the derivatives have been taken) we arrive at
\eqngrl{
\lim_{x\to y}b_n&=&{n!\ov (2n+1)!}\lap^{(n)}V}
{&-&\lim_{x\to y}\sum_{k=0}^{n-2}{n!\ov (2n-1-k)!}\lap_x^{(n-2-k)}
\sum_{p=0}^{n-2}\nabla_i b_p\nabla^i b_{k-p}.}{b9}
The terms proportional to $\nabla^ib\nabla_i b$ are at
least quadratic in the potential. Let us note that the terms
which are nonlinear in $V$ always
contain products of gradients (as $\nabla_iV\nabla^i V,\,
\nabla_i\lap V\nabla^iV$ etc.). The terms linear in $V$ in the
expansion (\ref{b6},\ref{b9}) correspond to the sum
of all terms of the forms $V,V^2,\dots,V\lap V,\lap^2 V$ etc. in
the Seeley-deWitt expansion. Thus in the linear approximation
one finds
\eqnl{
W(x;\tau)\equiv W(x,x;\tau)=\sum_{n=0}^\infty {n!\ov (2n+1)!}
\tau^n\lap^{(n)}V+O\big(\nabla V\cdot\nabla V\big).}{b10}
In one dimension or in the case when the potential $V$ depends
just on one variable, the series \refs{b10} can be converted
into the following integral
\eqnl{
W(x;\tau)=\sqrt{\pi\ov 4\tau}\int_{-\infty}^\infty V(y)
\Big[1-\Phi\big({\vert x-y\vert\ov \sqrt{\tau}}\big)\Big]dy
+O(\nabla V\cdot\nabla V),}{b11}
where $\Phi$ is the error function.
The nonlocal result \refs{b11} for $W$ accounts for all terms
which are linear in $V$. It is relevant for improving the
Coleman-Weinberg effective potential as well as the partition
function in statistical physics. It is related but not
identical to a similar expression obtained by Feynman
by variational method [15].


\begin{thebibliography}{99}
\bibitem{r1} S.W. Hawking, Commun. Math. Phys. {\bf 43} (1975) 199.
\bibitem{r2} V.P. Frolov and G.A. Vilkovisky, in
Proc. second seminar on quantum gravity (1981), Moscow, ed. M.A.
Markov and P.C. West, Plenum, London, 1983.
\bibitem{r3} A.O. Barvinsky and G.A. Vilkovisky, Nucl. Phys.
{\bf B333} (1990) 471.
\bibitem{r4} I.D. Novikov and V.P. Frolov, Physics of
Black Holes, Kluwer Acad. Publishers, Dordrecht/Boston/London, 1989.
\bibitem{r5} C.G. Callan, S.B. Giddings, J.A. Harvey and A. Strominger,
Phys. Rev. {\bf D45} (1992) 1005.
\bibitem{r6} L. Susskind and L. Thorlacius, Nucl. Phys. {\bf B382} (1992) 123.
\bibitem{r7} J.G. Russo, L. Susskind and L. Thorlacius, Phys. Rev.
{\bf D46} (1992) 3444.
\bibitem{r8} D.A. Love, Phys. Rev. {\bf D47} (1993) 2446.
\bibitem{r9} A.M. Polyakov, Phys. Lett. {\bf B103} (1981) 207.
\bibitem{r10} A.O. Barvinsky, Yu.V. Gusev, V.V. Zhutnikov and
G.A. Vilkovisky, preprint PRINT-93-0274, (1993) Manitoba.
\bibitem{r11} C.W. Misner, K.S. Thorne and J.A. Wheeler,
Gravitation, Freeman, San Francisco, 1973.
\bibitem{r12} J.S. Dowker and R. Critchley, Phys. Rev. {\bf D13} (1976)
3224; S.W. Hawking, Commun. Math. Phys. {\bf 55} (1977) 133.
\bibitem{r13} D.J. O'Connors, B.L. Hu and T.C. Shen, Phys. Lett.
{\bf 130B} (1983) 31; S. Blau, M. Visser and A. Wipf, Phys. Lett.
{\bf 209B} (1988) 209.
\bibitem{r14} N.D. Birrell and P.C.W.Davies,
Quantum Fields in Curved Space , Cambridge Univ. Press, 1982.
\bibitem{r15} R.P. Feynman, Statistical Mecchanics, W.A. Benjamin,
Massachusets, 1972.
\bibitem{r16} S. Coleman and E. Weinberg, Phys. Rev. {\bf D7} (1973) 1888.
\bibitem{r17} A.O. Barvinsky and G.A. Vilkovisky, Nucl. Phys.
{\bf B282} (1987) 163.
\end{thebibliography}
\end{document}